\documentclass[10pt,a4paper,twoside]{article}
\usepackage [latin1]{inputenc}
\usepackage{vmargin,fancyhdr,multicol,multirow,ifthen,cite,
            graphicx,wrapfig,calc,dcolumn,apalike,setspace,
            boxedminipage,rotating,textcomp,picinpar,longtable,
            url,amsmath,amssymb,ogonek,textgreek} 

\usepackage[tight]{subfigure}
\usepackage{imc2023}

\begin{document}
\SetPaperBodyFont
\setlength{\footskip}{3.60004pt}

\begin{IMCpaper}{
\title{A new approach to analyse single-site video observations: A Tau Her 2022 case study}
\author{N. O. Szab\'o$^{1,2}$, A. Igaz$^1$, 
M. R\'ozsahegyi$^1$, K. S\'arneczky$^1$, B. Cs\'ak$^1$, L. Deme$^1$, J. Vink\'o$^{1,2}$ and L. L. Kiss$^{1,2}$
        \thanks{$^1\,$Konkoly Observatory, CSFK Research Centre for Astronomy and Earth Sciences, \\
        Konkoly Thege \'ut 15-17, Budapest, 1121 Hungary \\
        $^2\,$ ELTE E\"otv\"os Lor\'and University, Institute of Physics, P\'azm\'any
    P\'eter s\'et\'any 1/A, Budapest, 1117, Hungary\\
                     \texttt{szabo.norton@csfk.org}}}

\abstract{Here we present a continuation of the work that was based on video observations of the Tau Her 2022 outburst from the McDonald Observatory, Texas, US. On the night of the maximum in 2022 we detected 626 individual Tau Her meteors, for which we estimated photovisual magnitudes and analysed their distribution on the sky to determine the radiant position in an innovative way. The derived population index is $5.56 \pm 1.83$, while the radiant position for the mid-time of the observations is: $\mathrm{RA} = 209.71 ^\circ \pm 1.01 ^\circ$, $\mathrm{Dec} = 27.73 ^\circ \pm 0.07 ^\circ$. Both measurements are in good agreement with other results in the literature, although our population index seems to be higher than most of the published values. We speculate this might be due to the higher sensitivity of our equipment to fainter meteors.}}%

\vspace*{-3\baselineskip}
\section{Introduction}

In 2022, the astronomical community witnessed the outburst of the Tau Herculids meteor shower (see, e.g. Hori et al. 2008, Rao  2021) predicted by Ye \& Vaubaillon (2022), an event that offered a unique opportunity for observational studies and analysis. Building upon our previous work presented at IMC 2022 (\cite{szaboetal2023}), where we reported on the initial findings from video observations at the McDonald Observatory, Texas, US, this paper delves deeper into the phenomena observed during this spectacular event.

Our earlier research focused on capturing and analyzing the Tau Herculids 2022 outburst, resulting in the detection of 626 individual meteors. Utilizing high-sensitivity video equipment (Slansky 2021), we were able to record and analyze these meteors, providing valuable insights into their distribution and characteristics. The primary objective was to estimate photovisual magnitudes and analyze their spatial distribution to innovatively determine the radiant position of the shower.

\begin{figure}[htb]
\centering
\includegraphics[width = \columnwidth]{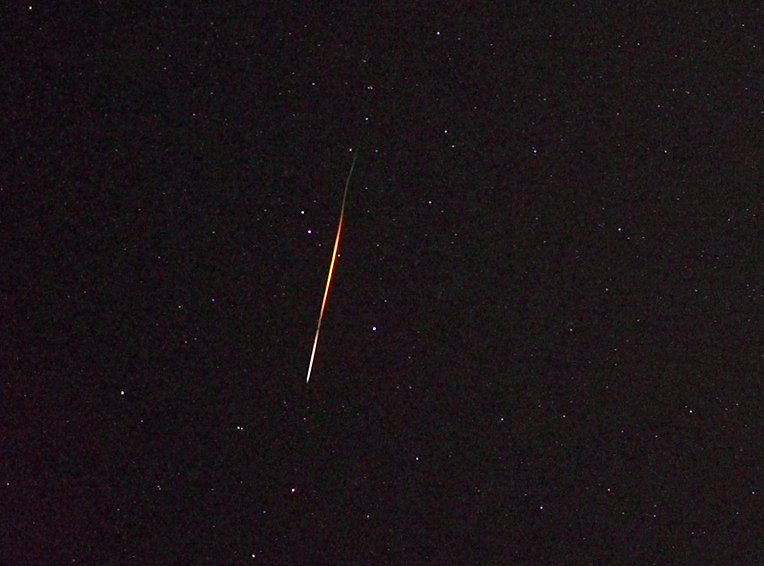}%
\vspace*{3pt}%
\caption{The brightest Tau Her meteor detected by us on the 31st of May 2022. We estimated its brightness to be -2 magnitudes as it passed through Corvus.}
\label{fig:nagy_meteor}
\end{figure}

This year, we present a continuation of our efforts, expanding upon the initial findings and offering a more comprehensive analysis of the Tau Herculids. Our observations have yielded a population index of $r = 5.56 \pm 1.83$, a figure that stands out when compared to other published values. This discrepancy, we hypothesize, could be attributed to the heightened sensitivity of our observational equipment, particularly adept at detecting fainter meteors. Furthermore, we have refined our measurements of the radiant position of the meteor shower for the mid-time of our observations.

\section{Data analysis}
\subsection{Population Index Determination}

The population index, $r$, is a key parameter that indicates the relative abundance of faint versus bright meteors in a shower. We estimated the brightness of each meteor during the Tau Herculids 2022 outburst visually, creating a distribution histogram on a logarithmic scale to better represent the wide range of meteor magnitudes.

Our histogram, as shown in Figure~\ref{fig:hist_meteor}, highlights a prominent decline in the number of bright meteors. This likely stems from a scarcity of bright comparison stars, which limited our ability to estimate the brightness for the most luminous meteors accurately. Additionally, the histogram shows a tapering at the faint end, suggesting a drop in detection efficiency for the dimmest meteors, possibly due to the limitations of our recording equipment.

\begin{figure}[htb]
\centering
\includegraphics[width = \columnwidth]{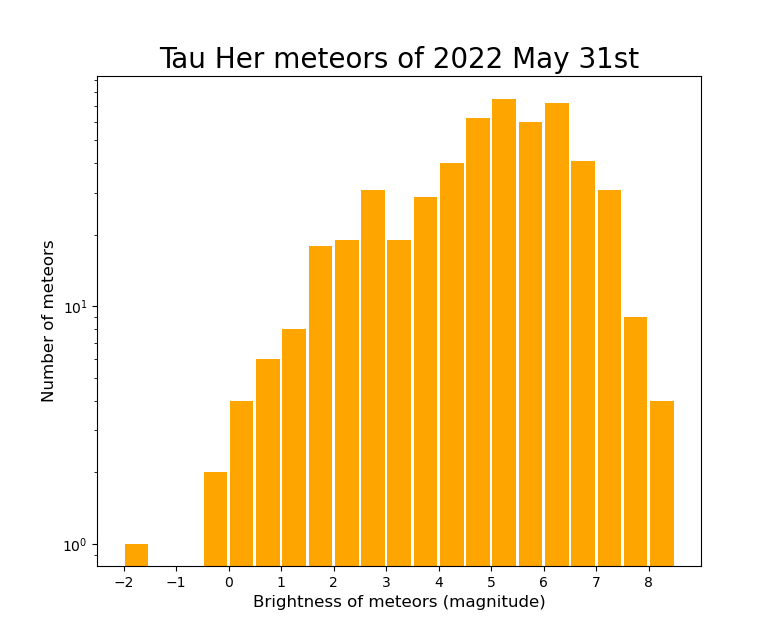}%
\vspace*{3pt}%
\caption{Logarithmic histogram showing the distribution of Tau Herculids meteor brightness. The x-axis represents meteor brightness in magnitudes, and the y-axis represents the number of meteors.}
\label{fig:hist_meteor}
\end{figure}

The derived population index from our data is $r = 5.56 \pm 1.83$, indicating a higher prevalence of faint meteors compared to bright ones. This value is notably higher than most published indices (Vauballion 2022, Rendtel \& Arlt 2022, Weiland 2022, Koten et al. 2022), suggesting that our high-sensitivity video equipment was particularly effective at detecting faint meteors.

\subsection{Radiant Position Calculation}

To determine the radiant of the meteor shower, we employed a method that circumvents the traditional pairing of meteor paths (Schmitt 2004). Each meteor path was projected onto a celestial sphere, and a great circle was fitted to its trajectory.

A heatmap of these great circle paths was generated to visualize the sky's density of meteor trajectories. The resulting heatmap, depicted in Figure~\ref{fig:radiant}, shows a clear concentration of paths that correspond to the radiant and anti-radiant of the shower. We then applied a two-dimensional Gaussian fit to the densest region to ascertain the precise coordinates of the radiant.

\begin{figure}[htbp]
\centering
\includegraphics[width=8cm]{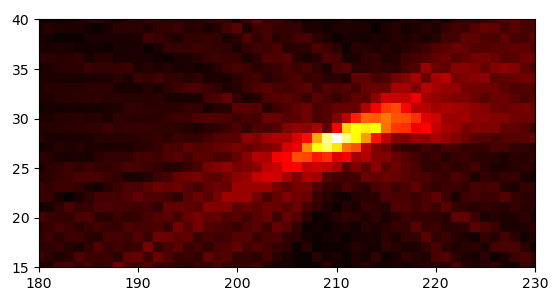}
\bigskip
\caption{Heatmap visualization of the sky, where the density of great circle paths of meteors is represented. The radiant location is indicated by the brightest areas.}
\label{fig:radiant}
\end{figure}

The coordinates for the radiant determined through our analysis are $\mathrm{RA} = 209.71^\circ \pm 1.01^\circ$, $\mathrm{Dec} = 27.73^\circ \pm 0.07^\circ$, providing a precise location for the mid-time of our observations. These coordinates align well with other results reported in the literature (Jenniskens 2022, Vida \& Segon 2022, Egal et al. 2022), reinforcing the validity of our methodologies and findings.

This method offers a scalable and efficient alternative to pairwise intersection, especially useful for large datasets.

Our findings present a detailed analysis of the Tau Herculids meteor shower, enhancing our understanding of its behavior and characteristics. The methods we've developed and employed have proven to be effective in handling the challenges associated with observing and analyzing meteor showers.

\section{Conclusions}
Our comprehensive study of the Tau Herculids meteor shower has provided valuable insights into the nature and behavior of this celestial phenomenon. Through meticulous observation and innovative analytical methods, we have not only reinforced findings from previous studies but also contributed new knowledge to the field of meteor astronomy.

The population index we derived is notably higher than the indices commonly reported in the literature. This suggests a significant presence of fainter meteors within the Tau Herculids stream, likely detected due to the superior sensitivity of our observational equipment. The implication of this finding is twofold: it underscores the importance of using high-sensitivity equipment in meteor observations to capture a more complete representation of the events, and it may also prompt a reevaluation of the population index in other meteor showers when similarly sensitive equipment is employed.

Furthermore, our determination of the radiant position provides a precise point of origin for the meteors observed during the peak of the Tau Herculids. The accuracy of these coordinates attests to the efficacy of our method, which utilizes a heatmap approach and a Gaussian fit, offering an innovative and scalable technique for radiant determination.

The findings from this study not only contribute to a deeper understanding of the Tau Herculids meteor shower but also demonstrate the potential for similar methodologies to be applied to other meteor showers, enhancing the accuracy and breadth of meteor observations. Future work will aim to refine these techniques further and possibly incorporate artificial intelligence for automated meteor detection and analysis.

\section*{Acknowledgements}

The authors are grateful for the financial support from the E\"otv\"os Lor\'and Research Network (project ID K\"O-31 "Kozmikus hat\'asok \'es kock\'azatok"). We also acknowledge generous support and assistance from Judit Gy\"orgyey-Ries, N\'ora Egei and Zolt\'an Bels\H{o}.

N. O. Sz. thanks the financial support provided by the undergraduate research assistant program of Konkoly Observatory.

\nocite{*}
\bibliographystyle{imo2}
\bibliography{ms}

\begin{thebibliography}{}

\bibitem[Egal et~al., 2023]{Egal_2023}
Egal A., Wiegert P.~A., Brown P.~G., and Vida D. (2023).
\newblock ``Modeling the 2022 \texttau-herculid outburst''.
\newblock {\em The Astrophysical Journal}, {\bf 949:2}, 96.

\bibitem[{Horii} et~al., 2008]{2008EM&P..102...85H}
{Horii} S., {Watanabe} J.-I., and {Sato} M. (2008).
\newblock ``{Meteor Showers Originated from 73P/Schwassmann Wachmann}''.
\newblock {\em Earth Moon and Planets}, {\bf 102:1-4}, 85--89.

\bibitem[{Jenniskens}, 2022]{jenniskens22}
{Jenniskens} P. (2022).
\newblock ``{Tau Herculid Meteors}''.
\newblock {\em Central Bureau for Astronomical Telegrams}, {\bf 5126}.

\bibitem[Koten et~al., 2023]{koten2023tau}
Koten P., Shrben{\`y} L., Spurn{\`y} P., Borovi{\v{c}}ka J., {\v{S}}tork R., Henych T., Voj{\'a}{\v{c}}ek V., and M{\'a}nek J. (2023).
\newblock ``Tau-herculid meteor shower on night 30/31 may, 2022, and properties of the meteoroids''.
\newblock {\em arXiv preprint arXiv:2305.13748}.

\bibitem[Quanzhi \& J{\'e}r{\'e}mie, 2022]{2022MNRAS.515L..45Y}
Quanzhi Y. and J{\'e}r{\'e}mie V. (2022).
\newblock ``{The 2022 encounter of the outburst material from comet 73P/Schwassmann-Wachmann 3}''.
\newblock {\em MNRAS}, {\bf 515:1}, L45--L49.

\bibitem[{Rao}, 2021]{rao21}
{Rao} J. (2021).
\newblock ``{Will Comet 73P/Schwassman-Wachmann 3 produce a meteor outburst in 2022?}''.
\newblock {\em WGN, Journal of the International Meteor Organization}, {\bf 49:1}, 3--14.

\bibitem[Rendtel \& Arlt, 2022]{rendtel2022tau}
Rendtel J. and Arlt R. (2022).
\newblock ``Tau herculids 2022: Rate, number density, population index and geometrical effects from visual data''.
\newblock {\em arXiv preprint arXiv:2209.12297}.

\bibitem[Schmitt, 2004]{schmidt}
Schmitt S. (2004).
\newblock ``Meteor shower radiant calculator''.
\newblock  {\tt http://www.convertalot.com/\\metior\_shower\_radiant\_calculator.html} .

\bibitem[{Slansky}, 2021]{2021JIMO...49..131S}
{Slansky} P.~C. (2021).
\newblock ``{The Fruits of Failure, Frustration and Fortune - Two years in an amateur meteor observer's life}''.
\newblock {\em WGN, Journal of the International Meteor Organization}, {\bf 49:5}, 131--133.

\bibitem[Szab\'o et~al., 2023]{szaboetal2023}
Szab\'o N., Igaz A., Kiss L., R\'ozsahegyi M., S\'arneczky K., Cs\'ak B., Deme L., and Vink\'o J. (2023).
\newblock ``International {M}eteor {C}onference 2022, {S}eptember 29--{O}ctober 2, {P}oroszl\'o, {Hungary}''.
\newblock {\em Proceedings of the IMC}, pages 120--123.

\bibitem[Vaubaillon et~al., 2023]{vaubaillon20232022}
Vaubaillon J., Loir C., Ciocan C., Kandeepan M., Millet M., Cassagne A., Lacassagne L., Da~Fonseca P., Zander F., Buttsworth D., et~al. (2023).
\newblock ``A 2022 $\tau$-herculid meteor cluster from an airborne experiment: automated detection, characterization, and consequences for meteoroids''.
\newblock {\em Astronomy \& Astrophysics}, {\bf 670}, A86.

\bibitem[{Vida} \& {Segon}, 2022]{vida22}
{Vida} D. and {Segon} D. (2022).
\newblock ``{Tau Herculid Meteors}''.
\newblock {\em Central Bureau for Astronomical Telegrams}, {\bf 5126}.

\bibitem[Weiland, 2022]{weiland2022analysis}
Weiland T. (2022).
\newblock ``Analysis of the unusual outburst of the tau-herculids in 2022, observed from {A}rizona, {USA}''.
\newblock {\em WGN, Journal of the International Meteor Organization}, {\bf 50:3-4}, 105--108.

\end{thebibliography}

\end{IMCpaper}
\end{document}